\documentclass[10pt,twocolumn,superscriptaddress,showpacs,preprintnumbers,amsmath,amssymb,prb,aps]{revtex4}
\usepackage{graphicx}
\usepackage{float}

\begin{document}


\title{Structure and superconductivity of two different phases of Re$_3$W}

\author{P.K. Biswas}
\email[]{P.K.Biswas@warwick.ac.uk}
\affiliation{Physics Department, University of Warwick, Coventry, CV4 7AL, United Kingdom}

\author{M.R. Lees}
\affiliation{Physics Department, University of Warwick, Coventry, CV4 7AL, United Kingdom}

\author{A.D. Hillier}
\affiliation{ISIS Facility, Science and Technology Facilities Council, Rutherford Appleton Laboratory, Chilton, Oxfordshire, OX11 0QX, U.K.}

\author{R.I. Smith}
\affiliation{ISIS Facility, Science and Technology Facilities Council, Rutherford Appleton Laboratory, Chilton, Oxfordshire, OX11 0QX, U.K.}

\author{W.G. Marshall}
\affiliation{ISIS Facility, Science and Technology Facilities Council, Rutherford Appleton Laboratory, Chilton, Oxfordshire, OX11 0QX, U.K.}

\author{D.McK. Paul}
\affiliation{Physics Department, University of Warwick, Coventry, CV4 7AL, United Kingdom}

\date{\today}

\begin{abstract}
Two superconducting phases of Re$_3$W have been found with different physical properties. One phase crystallizes in a non-centrosymmetric cubic ($\alpha$-Mn) structure and has a superconducting transition temperature, $T_{c}$, of 7.8~K. The other phase has a hexagonal centrosymmetric structure and is superconducting with a $T_{c}$ of 9.4~K. Switching between the two phases is possible by annealing the sample or remelting it. The properties of both phases of Re$_3$W have been characterized by powder neutron diffraction, magnetization, and resistivity measurements. The temperature dependence of the lower and the upper critical fields have been measured for both phases. These are used to determine the penetration depths and the coherence lengths for these systems.

\end{abstract}


\maketitle
\section{INTRODUCTION}
The discovery of superconductivity in the non-centrosymmetric (NCS) heavy fermion CePt$_3$Si (Ref.~\onlinecite{Bauer}) has resulted in a period of intense theoretical and experimental investigation into the physics of non-centrosymmetric superconducting materials. The lack of inversion symmetry in the crystal structure of this type of material along with strong spin-orbit (SO) coupling can lead to a mixing of spin-singlet and spin-triplet pair states~\cite{Gorkov}. These NCS materials exhibit unusual magnetic properties including suppressed paramagnetic limiting or high upper critical fields~\cite{Frigeri,Mineev} as seen in CePt$_3$Si,~\cite{Bauer} CeRhSi$_3$,~\cite{Kimura} and CeIrSi$_3$~\cite{Sugitani}, the appearance of superconductivity with antiferromagnet order in CePt$_3$Si,~\cite{Metoki}, and superconductivity at the border of ferromagnetism in UIr.~\cite{Akazawa}  
 
One recent focus of the work on non-centrosymmetric superconductors has been to investigate the properties of transition-metal compounds that have a significant spin-orbit coupling. Here, the complications of the $f$-electron heavy fermions, such as the strong electron correlations and the possibility of magnetically mediated superconductivity, are expected to be absent. The intermetallic Re$_3$W belongs in this category since it contains heavy atoms with atomic numbers 75 and 74 for Re and W respectively. Superconductivity in Re$_3$W was first reported in the 1960's. The material was shown to have a cubic $\alpha$-Mn structure,~\cite{Hulm,Blaugher} although it is worth noting that the authors of this early work did not comment on the fact that the $\alpha$-Mn structure is non-centrosymmetric. Since then, very little experimental work has been done on Re$_3$W. Recent penetration depth measurements carried out on the NCS phase of Re$_3$W by rf tunnel diode resonator and point-contact spectroscopy suggested that Re$_3$W is a weakly coupled isotropic s-wave superconductor.~\cite{Zuev,Kuznetsova,Huang}

In this paper, we report on the synthesis of two different superconducting phases of Re$_3$W. One phase has a centrosymmetric (CS) crystal structure, whereas the other has a non-centrosymmetric structure. Switching from the CS to the NCS phase is achieved by annealing the sample, while remelting the NCS sample in an arc furnace returns the sample to the CS structure. The ease with which one can switch between the two phases of Re$_3$W has allowed us to investigate and compare the properties of a CS and a NCS superconducting system using a single material without changes in stoichiometry. We characterize the properties of both phases of Re$_3$W using neutron diffraction, magnetization, $M$, and resistivity, $\rho$, measurements. We present the temperature dependence of the lower critical field, $H_{\rm{c1}}$, and the upper critical field, $H_{\rm{c2}}$, of both materials and also calculate the penetration depths and coherence lengths for these systems.

\section{SAMPLE PREPARATION}
Samples of the centrosymmetric phase of Re$_3$W were prepared  by melting together a stoichiometric mixture of Re lumps ($99.99\%$) and W pieces ($99.999\%$) in an arc furnace on a water-cooled copper hearth using tungsten electrodes in a high-purity Ar atmosphere. After the initial melt, the buttons were turned and remelted several times to ensure homogeneity. Samples of the non-centrosymmetric phase were made by annealing the as-grown samples for 5 days at $1500^\circ$C in a high-purity Ar atmosphere. The unannealed samples of Re$_3$W are hard but malleable. The samples become brittle after annealing. 

\begin{figure}[!htb]
\begin{center}
\includegraphics[width=1.0\columnwidth]{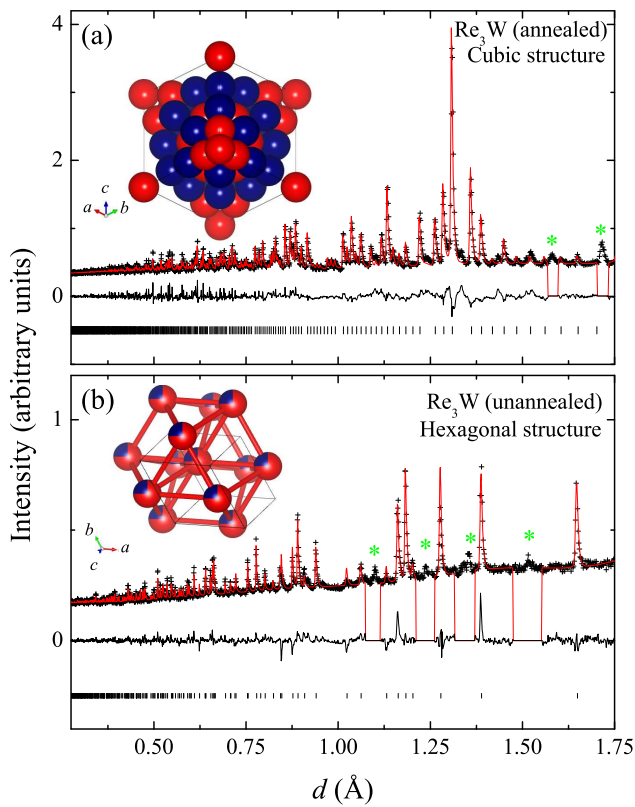}
\caption{\label{Figure1Biswas} (Color online) Neutron diffraction patterns as a function of $d$-spacing collected at 295~K for the annealed and unannealed samples of Re$_3$W. The experimental
data (crosses) are shown along with the calculated pattern obtained from a Rietveld refinement of the structure and a difference curve. The ticks indicate positions of the nuclear Bragg peaks. The green asterisks mark the positions of impurity peaks. The Rietveld refinement shows that the annealed sample has a non-centrosymmetric cubic structure while the unannealed sample has a centrosymmetric hexagonal structure. Insets show the crystal structures of the two phases, with the Re atoms shown in red (light gray) and the W atoms shown in blue (dark gray).}
\end{center}
\end{figure}
\section{RESULTS AND DISCUSSION}

\subsection{Powder neutron diffraction studies}
Time-of-flight powder neutron diffraction data were collected on the General Materials Diffractometer (GEM) at the ISIS Facility in the Rutherford Appleton Laboratory. UK~\cite{Day} The data were normalized to the incident neutron flux distribution and corrected for detector efficiencies and detector solid angle coverage, with the diffraction patterns collected in the highest resolution backscattering detector bank (average $2\theta=154^{\circ}$) used for structure refinement. Crystal structures were refined by the Rietveld method using the EXPGUI graphical interface for the GSAS package~\cite{Toby,Larson}. Figure~\ref{Figure1Biswas} shows the diffraction patterns collected at 295~K from the annealed and unannealed samples of Re$_3$W. The Rietveld refinement shows that the annealed sample has a cubic NCS ($\alpha$-Mn) structure (space group $I\bar{4}3m$) with a lattice parameter $a = 9.5987(7)$~\AA~[see Fig.~\ref{Figure1Biswas}(a)]. The diffraction pattern of the unannealed sample shows that this sample has a hexagonal structure with the CS space group $P6_3/mmc$ and lattice parameters $a = 2.770(3)$~\AA~ and $c = 4.5207(6)$~\AA~[see Fig.~\ref{Figure1Biswas}(b)]. The diffraction patterns of both phases of Re$_3$W contain some peaks (denoted by asterisks) that cannot be indexed, and these regions were excluded during the final stages of structure refinement. Insets of Figs.~\ref{Figure1Biswas}(a) and~\ref{Figure1Biswas}(b) show the crystal structures of the NCS and the CS phases of Re$_3$W. For the CS phase of Re$_3$W, both the Re and the W atoms share the same site leading to a random distribution of Re and W within the material. For the NCS phase, the refinement indicates that Re and the W atoms occupy preferred crystallographic sites and are therefore distributed in a more orderly fashion within the material. The refined composition of the NCS phase is not stoichiometric indicating there is still some uncertainty in the site occupation. Crystallographic parameters of the two phases of Re$_3$W are shown in Tables~\ref{table_of_refine} and \ref{table_of_atom}.

\begin{table}
\caption{Lattice parameters of the non-centrosymmetric and centrosymmetric phases of Re$_3$W determined from a structural refinement using the GSAS package of powder neutron diffraction data collected at 295~K.}
\label{table_of_refine}
\begin{center}
\begin{tabular}[b]{lll}\hline\hline
{}~~~~~~~~&NCS Re$_3$W~~~~~~~~&CS Re$_3$W\\\hline
Structure & Cubic & Hexagonal\\
Space group~~~~~~~~& $I\bar{4}3m$ & $P6_3/mmc$\\
\textit{a} (\AA) & 9.5987(7) & 2.770(3)\\
\textit{c} (\AA) &  & 4.5207(6)\\
$V_{\rm{cell}}$ ({\AA}$^3$) & 884.37(2) & 30.04(5)\\
$R_{\rm{p}}$ & 0.049 & 0.077\\
$wR_{\rm{p}}$ & 0.0697 & 0.1\\\hline\hline
\end{tabular}
\par\medskip\footnotesize
\end{center}
\end{table} 

\begin{table}
\caption{Atomic position parameters of the non-centrosymmetric and centrosymmetric phases of Re$_3$W determined from a structural refinement using the GSAS package of powder neutron diffraction data collected at 295~K}
\label{table_of_atom}
\begin{center}
\begin{tabular}[t]{lllllll}\hline\hline
NCS & Re$_3$W & & & & & \\\hline
Atom & Site& $x$ & $y$ & $z$ & Occ. & $U_{\rm{iso}}$ ({\AA}$^2$)\\\hline
Re & 2a & 0 & 0 & 0 & 0.99(4) & 0.070(5) \\
W & 8c & 0.3192(3) & 0.3192(3) & 0.3192(3) & 1.00 & 0.0068(7) \\
W & 24g & 0.3605(1) & 0.3605(1) & 0.0455(1) & 0.21(2) & 0.0090(3) \\
Re & 24g & 0.3605(1) & 0.3605(1) & 0.0455(1) & 0.78(1) & 0.0090(3) \\
Re & 24g & 0.0910(1) & 0.0910(1) & 0.2825(1) & 1.00 & 0.0089(2) \\\hline\hline
CS & Re$_3$W & & & & & \\\hline
Atom & & $x$ & $y$ & $z$ & Occ. & $U_{\rm{iso}}$ ({\AA}$^2$)\\\hline
Re & 2c & 0.3333 & 0.6667 & 0.25 & 0.75 & 0.0025(1) \\
W & 2c & 0.3333 & 0.6667 & 0.25 & 0.25 & 0.0025(1) \\\hline\hline
\end{tabular}
\par\medskip\footnotesize
\end{center}
\end{table} 

\begin{figure}[!htb]
\begin{center}
\includegraphics[width=1.0\columnwidth]{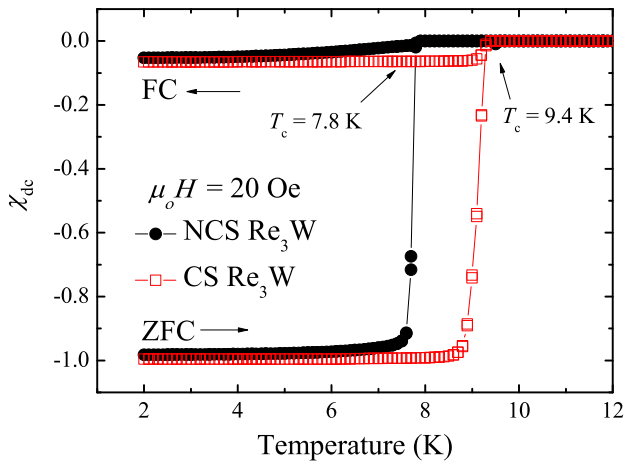}
\caption{\label{Figure2Biswas} (Color online) Temperature dependence of the magnetic susceptibility for the non-centrosymmetric and the centrosymmetric Re$_3$W measured in zero-field-cooled and field-cooled mode in an applied magnetic field of 20~Oe.}
\end{center}
\end{figure}

\subsection{Magnetization and resistivity measurements}
Magnetization measurements were made as a function temperature in an applied magnetic field of 20~Oe using a Quantum Design Magnetic Property Measurement System (MPMS) magnetometer. The temperature dependence of the dc magnetic susceptibility, $\chi_{\rm{dc}}(T)$, shows that the NCS Re$_3$W sample has a superconducting transition temperature, $T_{c}^{\rm{onset}}$, of 7.8~K [see Fig.~\ref{Figure2Biswas}] with a transition width $\Delta{T}_{c}=0.21$~K. For CS Re$_3$W, the onset of the transition is around 9.4~K with a much broader transition of $\Delta{T}_{c}=0.50$~K. Comparable transition widths are observed in the resistivity measurements (see below). This suggests, as expected, that the annealed NCS phase of Re$_3$W is more ordered than the unannealed CS phase. At 2~K, the zero-field-cooled (ZFC) dc susceptibility approaches a value of -1 ($\sim{100\%}$ shielding effect) for both the samples, while the field-cooled (FC) signal shows a flux exclusion (Meissner effect) of $\sim{5\%}$ for the NCS phase and $\sim{7\%}$ for the CS phase.

\begin{figure}[!htb]
\begin{center}
\includegraphics[width=1.0\columnwidth]{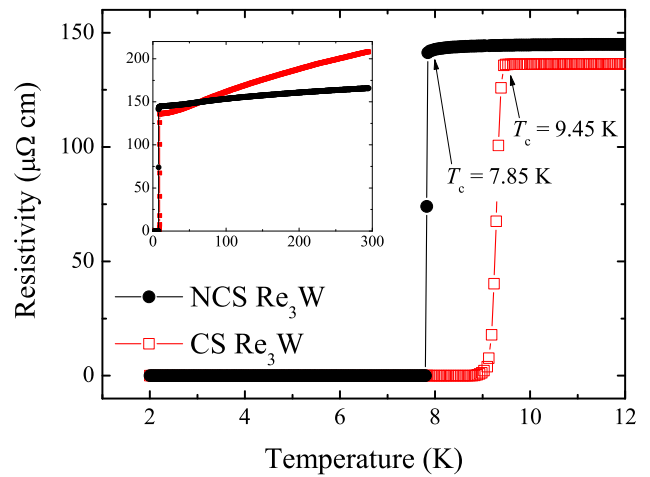}
\caption{\label{Figure3Biswas} (Color online) Low-temperature ac electrical resistivity of the NCS and the CS phases of Re$_3$W. The inset shows the electrical resistivity of both phases up to room temperature.}
\end{center}
\end{figure}

The ac electrical resistivity was measured as a function of temperature, $\rho(T)$, for both phases of Re$_3$W via a standard four-probe method using a Quantum Design Physical Property Measurement System (PPMS) [see Fig.~\ref{Figure3Biswas}]. The NCS Re$_3$W shows a superconducting transition (onset) at 7.85~K ($\Delta{T}_{c}=0.05$~K) while the CS Re$_3$W has a transition at 9.45~K ($\Delta{T}_{c}=0.32$~K). The zero-field onset transition temperatures determined from the resistivity measurements are slightly higher than those obtained from the magnetization measurements performed in 20 Oe. The resistivity curves between 2 and 295~K show metallic behavior for the CS phase of Re$_3$W, whereas it is almost temperature independent above $T_{c}$ in the NCS phase of Re$_3$W [see the inset of Fig.~\ref{Figure3Biswas}]. The relative resistance ratio, $\rho(295~K)/\rho(10~K)$ and the room temperature resistivity are 1.15(1) and $1.7\mu\Omega${m} for the NCS phase and 1.52(1) and $2.1\mu\Omega${m} for the CS phase, indicating that both samples are poor metals. The NCS phase is the more brittle of the two materials and any extrinsic factors such as microscopic cracks in the sample are more likely to play a role in this material. Given that the room temperature resistivity of the NCS sample is lower than the CS phase we suggest that cracks are not the reason for the high normal-state resistivity. The poor conductivity is more likely to result from a combination of strong electronic scattering and a large temperature independent residual resistivity due to structural disorder (while the NCS annealed phase is structurally more ordered than the CS phase, the NCS phase still retains a degree of Re/W disorder). We have calculated the mean free path, $l_{tr}$, based on the BCS approach~\cite{Orlando} using the room temperature resistivity data. The calculations yield $l^{NCS}_{tr}=0.277$~nm~and $l^{CS}_{tr} =0.224$~nm. These values are comparable with size of the crystallographic unit cells and given the coherence lengths, $\xi$, derived below, indicate that both phases of Re$_3$W are in the dirty limit.

\begin{figure}[!htb]
\begin{center}
\includegraphics[width=1.0\columnwidth]{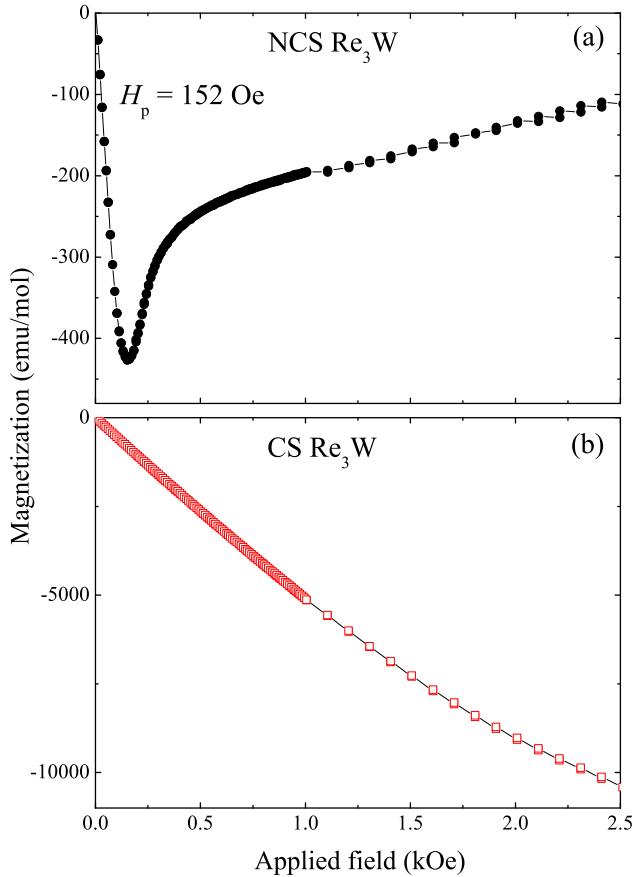}
\caption{\label{Figure4Biswas} (Color online) Virgin magnetization curves measured at 1.8~K for the (a) non-centrosymmetric and (b) centrosymmetric phases of Re$_3$W.}
\end{center}
\end{figure}

\begin{figure}[!htb]
\begin{center}
\includegraphics[width=1.0\columnwidth]{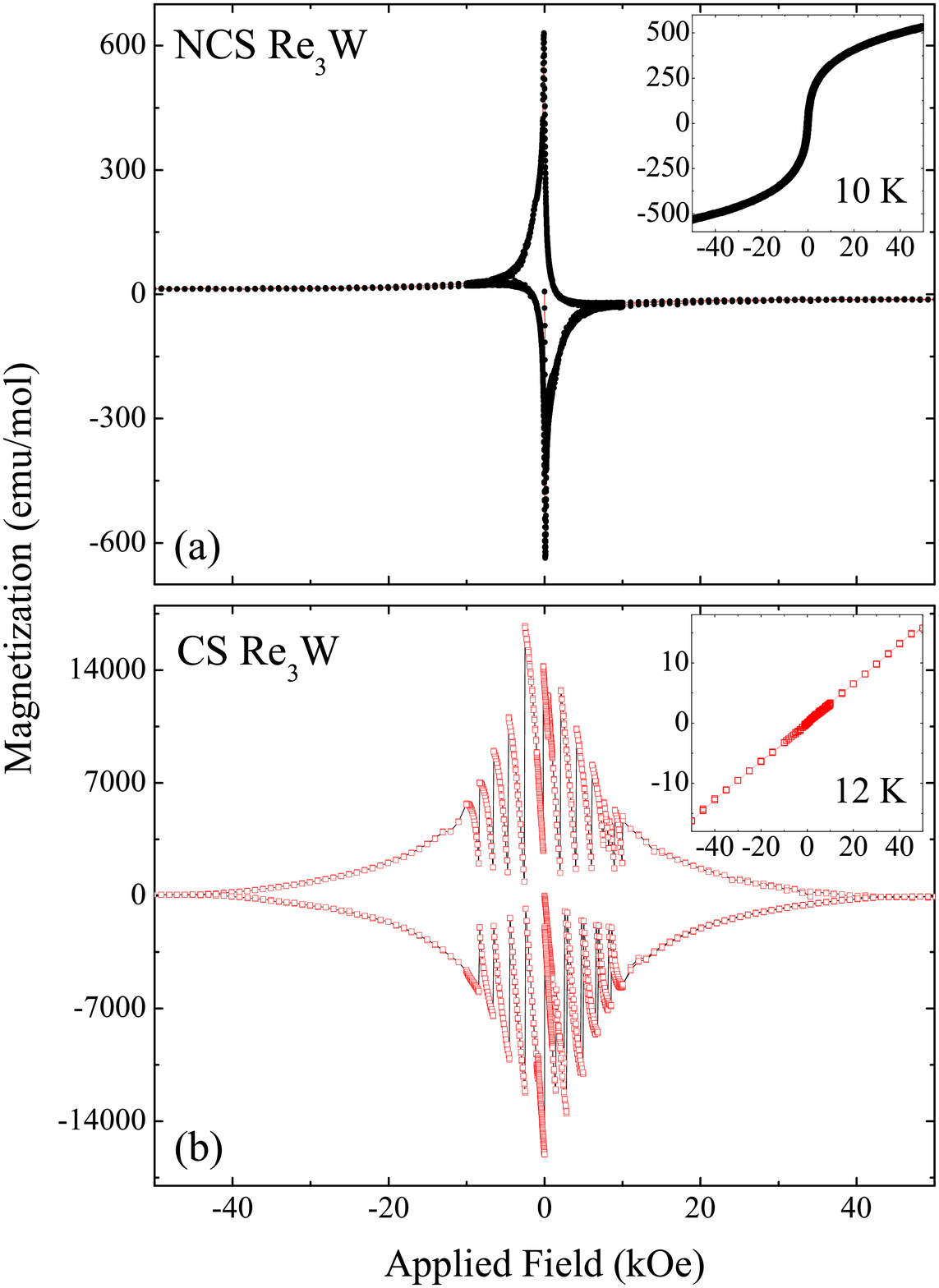}
\caption{\label{Figure5Biswas} (Color online) Magnetization hysteresis loops at 1.8~K for the (a) NCS and (b) CS phases of Re$_3$W. The insets show the $M(H)$ curves for the samples above $T_c$.}
\end{center}
\end{figure}

Figs.~\ref{Figure4Biswas}(a) and (b) show the low-field virgin magnetization data for the NCS and CS phases of Re$_3$W at 1.8~K and figures~\ref{Figure5Biswas}(a) and (b) show the full magnetization versus applied magnetic field loops collected in the superconducting state at 1.8~K. For the NCS sample the raw $M(H)$ data contain a significant paramagnetic contribution. This contribution has been removed from the data shown in Fig.~\ref{Figure5Biswas}(a) by measuring an $M(H)$ curve above $T_{c}$ at 10~K (see inset of Fig.~\ref{Figure5Biswas}(a)). The signal for the CS sample contains a small linear susceptibility  $\chi_{\rm{dc}}=3.53\times10^{-3}$~emu/mol (see inset of Fig.~\ref{Figure5Biswas}(b)) that has been subtracted from the data shown in figure~\ref{Figure5Biswas}(b). For the NCS sample, the value of $H_{\rm{p}}$ (complete penetration of the magnetic field) is 152~Oe while in the CS sample, $H_{\rm{p}}> 2500$~Oe. For the NCS phase of Re$_3$W the magnetization is reversible all the way from 70~kOe, the highest field that we can apply in our magnetometer, down to 10~kOe. We presume that the magnetization will remain reversible up to the upper critical field, estimated from the resistance measurements presented below to be 113~kOe at $T=1.8$~K. In contrast, the magnetization of the CS sample only becomes reversible in magnetic fields above 40~kOe (with $H_{\rm{c2}}(T=1.8~\rm{K})\approx130$~kOe). For lower fields the hysteresis loops of the CS sample contain a number of large magnetic-flux jumps, while no jumps are observed for the annealed samples. These flux jumps occur at lower temperatures ($T\leq4$~K) and at applied fields below $\sim20$~kOe. The number and magnitude of the flux jumps vary from loop to loop and become less frequent as the field sweeping rate, $dH/dt$, is decreased (data not shown). At 5~K with $(dH/dt\leq10$~Oe/s) no flux jumps are observed and at higher temperatures (5~K~$<T<T_{c}$) the flux jumps disappear.

The $M(H)$ curves show that both Re$_3$W phases exhibit reversible behavior below $T_{c}$ over a large region of the $H$-$T$ phase diagram. These data indicate that the bulk pinning is stronger in the CS phase of Re$_3$W than the NCS phase, and that the flux jumps are due to thermomagnetic instabilities induced by the motion of vortices into the superconductor combined with the sudden redistribution of the vortices within the sample~\cite{Mints}. The symmetry of the loops suggests that surface barriers do not play an important role in determining the form of the magnetization loops in this material. Further studies are underway to investigate the different pinning mechanisms in the two different phases of Re$_3$W.

\begin{figure}[!htb]
\begin{center}
\includegraphics[width=1.0\columnwidth]{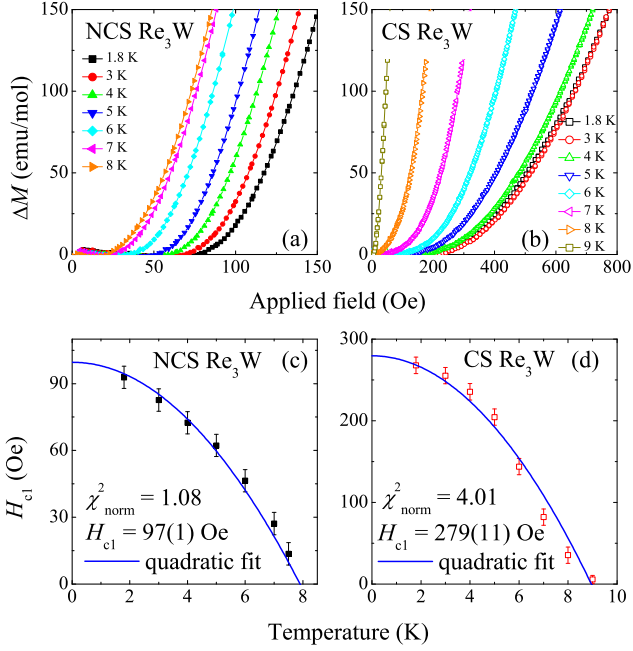}
\caption{\label{Figure6Biswas} (Color online) Deviation, $\Delta M$, from the linear virgin magnetization as a function of applied field determined at different temperatures for the (a) NCS and (b) CS phases of Re$_3$W. $H_{\rm{c1}}$ as a function of temperature for the (c) NCS and (d) CS phases of Re$_3$W. The solid lines are fits to the data using the expression $H_{c1}(T)=H_{c1}(0)\left\{1-(T/T_c)^2\right\}$.}
\end{center}
\end{figure}

The value of the lower critical field, $H_{\rm{c1}}$, was determined by measuring the field of first deviation from the initial slope of the magnetization curve. To this end, a linear fit to the data between 0 and 10~Oe was made. The deviation from linearity, $\Delta{M}$, was then calculated by subtracting this fit from the magnetization curves and plotted as a function of applied field [see Figs.~\ref{Figure6Biswas}(a) and ~\ref{Figure6Biswas}(b)]. The temperature dependence of $H_{\rm{c1}}$ for the two phases of Re$_3$W are obtained by using the criterion $\Delta{M}={10}^{-4}$~emu/mol and plotted in Figs.~\ref{Figure6Biswas}(c) and ~\ref{Figure6Biswas}(d). Demagnetizing effects are taken into account in estimating the $H_{\rm{c1}}$ values. The values determined for both the samples are fitted using the expression $H_{c1}(T)=H_{c1}(0)\left\{1-(T/T_c)^2\right\}$, where $H_{\rm{c1}}(0)$ is the lower critical field at zero temperature. The quadratic equation fits the data well for the NCS phase, whereas the model provides a poor fit to the data of the CS phase. The fits yield $H_{\rm{c1}}(0)$ of 97(1) and 279(11)~Oe for the NCS and the CS samples respectively.

\begin{figure}[!htb]
\begin{center}
\includegraphics[width=1.0\columnwidth]{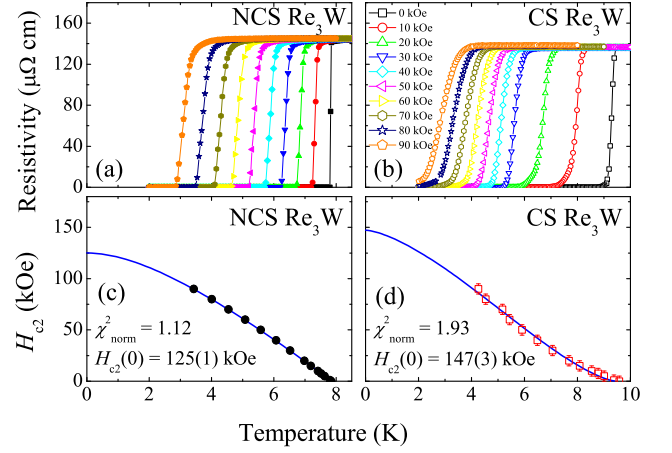}
\caption{\label{Figure7Biswas} (Color online) Temperature variation of the resistivity in a set of magnetic fields from 0 to 90~kOe for the (a) NCS and (b) CS phases of Re$_3$W. Temperature dependence of the upper critical fields of the (c) NCS and (d) CS phases of Re$_3$W. The solid line in (c) is a fit to the data using the WHH model. The solid line in (d) is a fit to the data using $H_{c2}(T)=H_{c2}(0)(1-(T/T_{c})^{3/2})^{3/2}$.}
\end{center}
\end{figure}

Resistivity measurements as a function of temperature were made in magnetic fields up to 90~kOe on both the NCS and the CS samples of Re$_3$W  [see Fig.~\ref{Figure7Biswas} (a) and (b)]. In the normal state just above $T_{c}$, both phases exhibit a small ($\sim0.6$\%) positive magnetoresistance in a magnetic field of 90~kOe. The temperature dependence of the upper critical field, $H_{c2}(T)$, of the NCS and the CS samples, determined from the onset of the resistive transitions (defined by a 1\% drop of the resistivity), are shown in Fig.~\ref{Figure7Biswas} (c) and (d). For the NCS sample, the temperature dependence of $H_{c2}$ is nearly linear close to $T_{c}$ with $dH^{\rm{NCS}}_{c2}/dT=-23.0(2)$~kOe/K and can be described using the Werthamer-Helfand-Hohenberg (WHH) model~\cite{Werthamer,Helfand}. In fitting the data to the WHH relations we calculated the Maki parameter, $\alpha=1.41$, compared with a value of 1.22 estimated from the gradient $dH^{\rm{NCS}}_{c2}/dT$ near $T_{c}$. The WHH fit yields $H^{\rm{NCS}}_{c2}=125(1)$~kOe at $T=0$~K. The temperature dependence of $H_{c2}$ for the CS sample clearly shows a difference in behavior compared to the NCS sample with deviations from the conventional WHH dependence. The data have a positive curvature with temperature near $T_{c}$ and are linear thereafter. A similar behavior is observed in polycrystalline borocarbides~\cite{Shulga}, MgB$_2$~\cite{Shigeta,Takano}, and Nb$_{0.18}$Re$_{0.82}.$~\cite{Karki} A reasonable fit to the data for the CS sample can be obtained using the expression $H_{c2}(T)=H_{c2}(0)(1-(T/T_{c})^{3/2})^{3/2}$ giving $H^{\rm{CS}}_{c2}(0)=147(3)$~kOe~\cite{Micnas}. A simple linear extrapolation of the lower temperature data to $T=0$~K gives $H^{\rm{CS}}_{c2}(0)=178(5)$~kOe with $dH^{\rm{CS}}_{c2}/dT=-21(1)$~kOe/K. A similar analysis of the data for the NCS sample gives $H^{\rm{NCS}}_{c2}(0)=153(1)$~kOe. While the temperature dependence of $H_{c2}$ for the two phases is clearly different, the values of $H_{c2}$ at $T=0$~K are comparable. The analysis presented above shows that $H_{c2}(0)$ appears to be slightly higher in the CS phase. Measurements at higher fields and lower temperatures are required to reveal to what extent Pauli limiting plays a role in determining $H_{c2}(0)$ in these materials. 

The coherence length, $\xi$, can be calculated using the Ginzburg-Landau (GL) relation $\xi=(\Phi_{\circ}/2\pi{H}_{c2})^{1/2}$, where $\Phi_{\circ}= 2.068 \times10^{-15}$~Wb is the flux quantum~\cite{Brandt}. With $H^{\rm{NCS}}_{c2}(0)=125(1)$~kOe for NCS Re$_3$W, the estimated $\xi^{\rm{NCS}}(0)$ is 5.13(2)~nm. For CS Re$_3$W, the value of  $\xi^{\rm{CS}}(0)$ is deduced to be 4.73(5)~nm from $H^{\rm{CS}}_{c2}(0)=147(3)$~kOe. Combining $\xi$ and the standard expression $H_{c1}=\frac{\Phi_{0}}{4\pi\lambda^2}\left(\ln\frac{\lambda}{\xi}+0.12\right),$~\cite{Tinkham} we estimate the magnetic penetration depth, $\lambda^{\rm{NCS}}(0)=257(1)$~nm and $\lambda^{\rm{CS}}(0)=141(11)$~nm for the NCS and CS phases of Re$_3$W, respectively. The value of $\lambda^{\rm{NCS}}(0)$ is in good agreement with the value of 300(10)~nm reported by Zuev \textit{et al}.~\cite{Zuev}. We used the values of $\lambda(0)$ and $\xi(0)$ to calculate the GL parameter $\kappa=\lambda/\xi$. They yield $\kappa^{\rm{NCS}}(0)=50(1)$ for the NCS phase and $\kappa^{\rm{CS}}=30(3)$ for the CS phase of Re$_3$W. The measured and derived superconducting and transport parameters of the NCS and the CS phases of Re$_3$W are listed in Table~\ref{table_of_fits}.

\begin{table}
\caption{Measured and derived superconducting and transport parameters of the non-centrosymmetric and centrosymmetric phases of Re$_3$W.}
\label{table_of_fits}
\begin{center}
\begin{tabular}[t]{lll}\hline\hline
{}~~~~~~~~&NCS Re$_3$W~~~~~~~~&CS Re$_3$W\\\hline
$T_{c}^{\rm{onset}}$ (K) & 7.80$\pm$0.05 & 9.40$\pm$0.05\\
$H_{\rm{c1}}(0)$ (Oe) & 97$\pm$1 & 279$\pm$11\\
$H_{c2}(0)$ (kOe)~~~~~~~~& 125$\pm$1 & 147$\pm$3\\
$\lambda(0)$ (nm) & 257$\pm$1 & 141$\pm$11\\
$\xi(0)$ (nm) & 5.13$\pm$0.02 & 4.73$\pm$0.05\\
$\kappa(0)$ & 50$\pm$1 & 30$\pm$3\\
$\rho$(295~K)($\mu\Omega$m) & 1.7 & 2.1\\
$l_{tr}$~(nm) & 0.277 & 0.224\\\hline\hline
\end{tabular}
\par\medskip\footnotesize
\end{center}
\end{table} 

\section{SUMMARY}
In summary, we have shown that there are two different phases of Re$_3$W. One phase is non-centrosymmetric with an $\alpha$-Mn structure and is superconducting with a $T_{c}$ of 7.8~K. The other phase is centrosymmetric with a hexagonal structure and is also superconducting with a $T_{c}$ of 9.4~K. Switching between the two phases is made possible by annealing (CS to NCS) or remelting (NCS to CS) the samples. The full hysteresis loops of the CS sample of Re$_3$W show giant flux jumps, while no jumps are observed for the NCS sample. The temperature dependence of $H_{c2}$ of the NCS phase for Re$_3$W can be fitted using the WHH model which yields $H^{\rm{NCS}}_{c2}(0)=125(1)$~kOe. In contrast, $H_{c2}(T)$ of the CS phase of Re$_3$W is linear at lower temperature and has a positive curvature nearer to $T_{c}$. A fit to the data gives $H^{\rm{CS}}_{c2}(0)=147(3)$~kOe. Using GL relations, the penetration depths are estimated to be $\lambda^{\rm{NCS}}=257(1)$~nm~and $\lambda^{\rm{CS}}=141(11)$~nm~and the coherence lengths are calculated to be $\xi^{\rm{NCS}}=5.13(1)$~nm~and $\xi^{\rm{CS}}=4.73(1)$~nm at $T=0$~K. Our results compare well with unpublished work~\cite{Kuznetsova} on the NCS phase of Re$_3$W. To the best of our knowledge, Re$_3$W is the first system of its kind to be found to exist as either a centrosymmetric or a non-centrosymmetric superconducting material. This, and similar systems if they exist, offer a good opportunity to study the interplay between the structure, spin-orbit coupling, and superconducting properties of intermetallic systems.

\begin{acknowledgments}
This work was supported by the Engineering and Physical Sciences Research Council (EPSRC) and the Science and Technology Facilities Council (STFC). Xpress Access neutron beam time on GEM was provided by the UK Science and Technology Facilities Council (STFC). PKB would like to thank the Midlands Physics Alliance Graduate School (MPAGS) for a studentship. The Quantum Design MPMS magnetometer and PPMS used in this research were obtained through the Science City Advanced Materials project: Creating and Characterizing Next Generation Advanced Materials project, with support from Advantage West Midlands (AWM) and part funded by the European Regional Development Fund (ERDF).
\end{acknowledgments}

\bibliography{BiswasRevised}

\end{document}